\documentclass[twocolumn]{aastex61}
\usepackage{hyperref}
\usepackage{amsmath,amstext}
\usepackage[T1]{fontenc}
\usepackage[figure,figure*]{hypcap}
\usepackage{graphicx}
\usepackage{multirow}

\def\sun{$_{\odot}$}

\def\f28{${f}_{2-8{\rm keV}}$}

\def\sun{$_{\odot}$}

\def\ergscm2{erg s$^{-1}$ cm$^{-2}$}
\def\yr-1{yr$^{-1}$}

\def\asec{\ifmmode^{\prime\prime}\else$^{\prime\prime}$\fi}

\def\secspt{$\buildrel{\prime\prime}\over .$}

\def\Chandra{{\it Chandra}}

\received{December 7, 2017}
\submitjournal{ApJ Letters}

\shorttitle{Brightening X-rays From GW170817/GRB170817A} 
\shortauthors{Ruan {\it et al.}}

\begin{document}

\title{Brightening X-ray Emission from GW170817/GRB170817A: \\ Further Evidence for an Outflow}

\correspondingauthor{John J. Ruan}
\email{john.ruan@mcgill.ca}

\author[0000-0001-8665-5523]{John J.~Ruan}
\affil{McGill Space Institute and Department of Physics, McGill University, 3600 rue University, Montreal, Quebec, H3A 2T8, Canada}

\author[0000-0002-3310-1946]{Melania Nynka}
\affil{McGill Space Institute and Department of Physics, McGill University, 3600 rue University, Montreal, Quebec, H3A 2T8, Canada}

\author[0000-0001-6803-2138]{Daryl Haggard}
\affil{McGill Space Institute and Department of Physics, McGill University, 3600 rue University, Montreal, Quebec, H3A 2T8, Canada}
\affil{CIFAR Azrieli Global Scholar, Gravity \& the Extreme Universe Program, Canadian Institute for Advanced Research, 661 University Avenue,
Suite 505, Toronto, ON M5G 1M1, Canada}

\author[0000-0001-6803-2138]{Vicky Kalogera}
\affil{Center for Interdisciplinary Exploration and Research in Astrophysics and Department of Physics and Astronomy, Northwestern University, 2145 Sheridan Road, Evanston, Illinois 60208-3112, USA}

\author[0000-0002-8465-3353]{Phil Evans}
\affil{Leicester Institute for Space and Earth Observation and Department of Physics \& Astronomy, University of Leicester, University Road, Leicester, LE1 7RH, UK}

\begin{abstract}
The origin of the X-ray emission from neutron star coalescence GW170817/GRB170817A is a key diagnostic of the unsettled post-merger narrative, and different scenarios predict distinct evolution in its X-ray light curve. Due to its sky proximity to the Sun, sensitive X-ray monitoring of GW170817/GRB170817A has not been possible since a previous detection at 16 days post-burst. We present new, deep \Chandra\ observations of GW170817/GRB170817A at 109 days post-burst, immediately after Sun constraints were lifted. The X-ray emission has brightened from a 0.3-8.0 keV flux of $3.6\times10^{-15}$ erg s$^{-1}$ cm$^{-2}$ at 16 days to $15.8\times10^{-15}$ erg s$^{-1}$ cm$^{-2}$ at 109 days, at a rate similar to the radio observations. This confirms that the X-ray and radio emission have a common origin. We show that the X-ray light curve is consistent with models of outflow afterglows, in which the outflow can be a cocoon shocked by the jet, dynamical ejecta from the merger, or an off-axis structured jet. Further deep X-ray monitoring can place powerful constraints on the physical parameters of these models, by both timing the passing of a synchrotron cooling break through the X-ray band, and detecting the associated steepening of the X-ray photon index. Finally, the X-ray brightening strengthens the argument that simple off-axis top-hat jet models are not consistent with the latest observations of GW170817/GRB170817A. 
\end{abstract}

\section{Introduction}
\label{sec:intro}

The gravitational wave (GW) and multi-wavelength electromagnetic (EM) discoveries of the binary neutron star (NS) merger GW170817 marked the dawn of multi-messenger GW astronomy \citep[e.g.,][]{abbott17a, abbott17b, coulter17, evans17, goldstein17, hallinan17, soares17, troja17, valenti17}. Detection of the short Gamma-ray burst (sGRB) GRB170817A associated with the gravitational wave event GW170817 confirmed that binary NS mergers are the progenitors of at least some sGRBs \citep{abbott17c, goldstein17, savchenko17}. Furthermore, observations of the optical and infrared transient confirmed that binary NS mergers produce kilonova, powered by r-process nucleosythesis \citep{arcavi17, cowperthwaite17, drout17, kasliwal17, mccully17, pian17, shappee17, smartt17}. Finally, robust association of GW170817 with its host galaxy \citep{hjorth17, im17, levan17} enabled measurements of the Hubble constant independent of the cosmic distance ladder \citep{abbott17d, guidorzi17}, providing a new probe of cosmology. However, many questions about the nature of GW170817 remain open.

\begin{figure*} [t!]
\center{
\includegraphics[scale=0.25,angle=0]{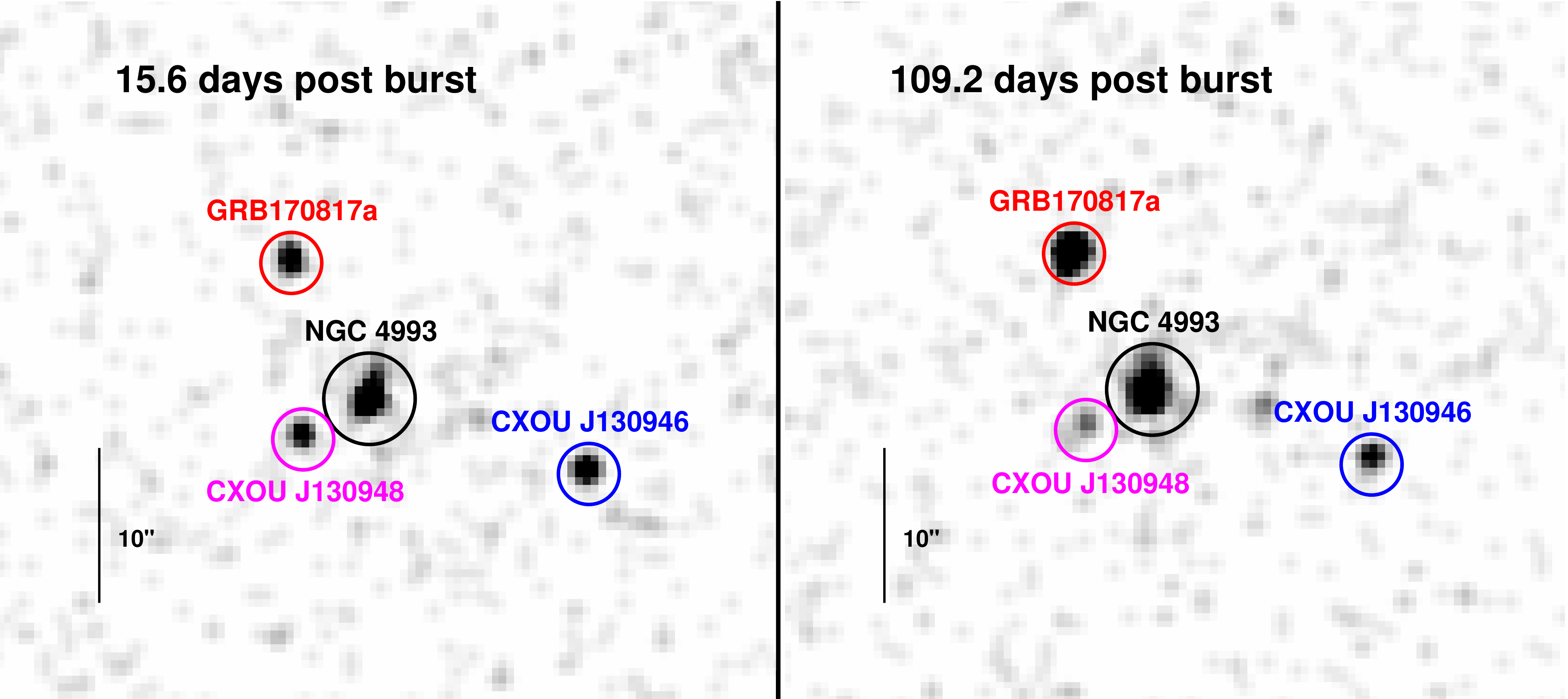}
\includegraphics[scale=0.344, angle=0.0]{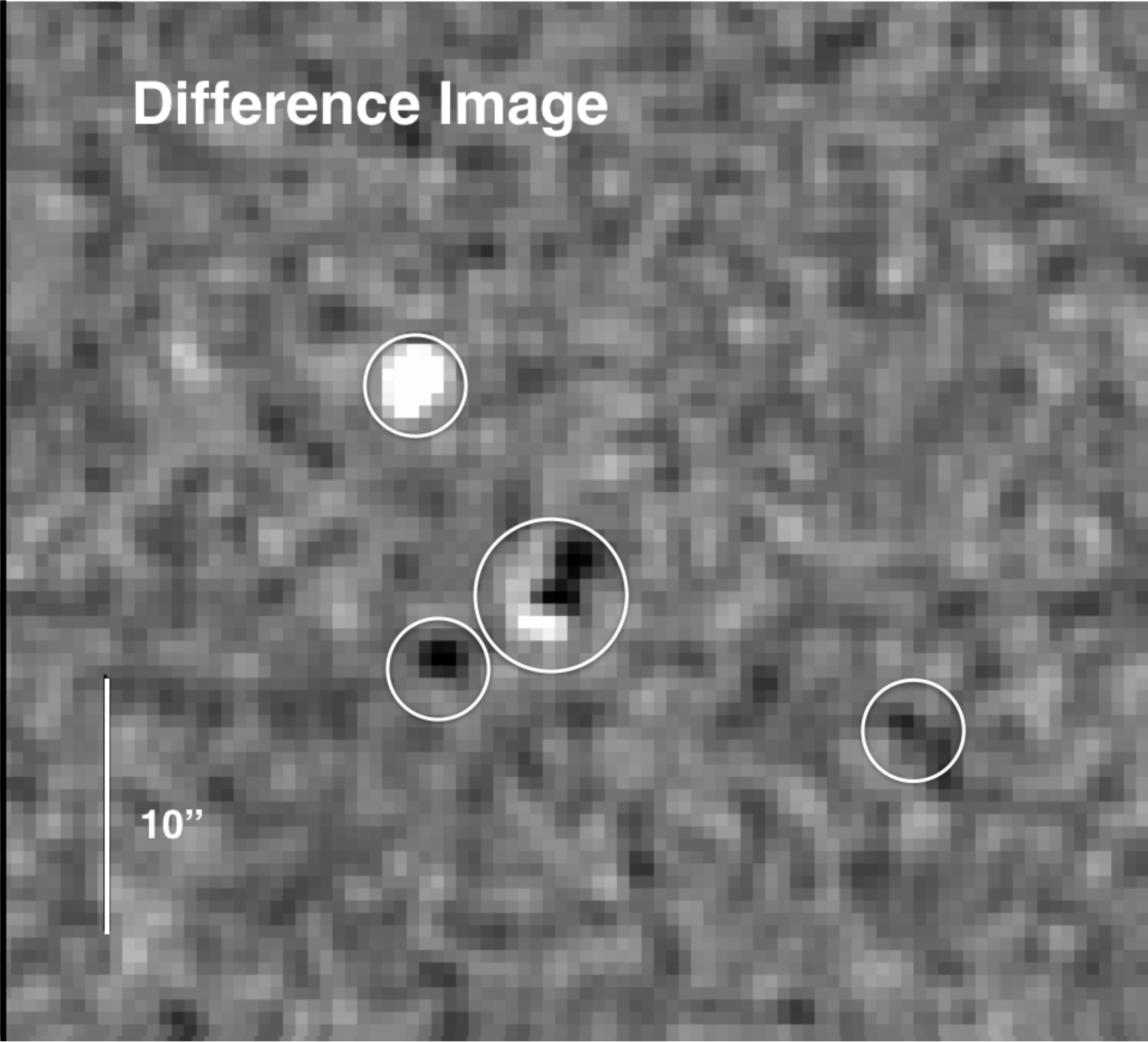}
}
\figcaption{\textit{\textbf{Left}}: \Chandra\ $0.5-8.0$ keV X-ray image of GRB170817A at 15.6 days post-burst, in a 93.4~ks observation from \citet{haggard17}. X-ray emission from GRB170817A is clearly detected, along with the host-galaxy NGC~4993 and two other sources in the field. \Chandra\ observations beyond this observations were Sun-constrained until early December 2017. 
\textit{\textbf{Middle}}: A subsequent 98.8~ks \Chandra\ image at $\sim$109.2 days post-burst, immediately after Sun constraints were lifted. The X-ray emission from GRB170817A is still detected and has brightened. 
\textit{\textbf{Right}}: A difference image in which the 15.6 day image is subtracted from the 109.2 day image, scaled by their respective exposures.  GRB170817A is clearly brightening, as indicated by its excess emission (white), while the emission deficit (black) from the variable source CXOU~J130948 indicates its decrease in flux. All images are shown on a linear scale, smoothed with a 2-pixel Gaussian kernel.}
\label{fig:image}
\end{figure*}

\begin{figure*}[t!]
\center{
\includegraphics[scale=0.32,angle=0]{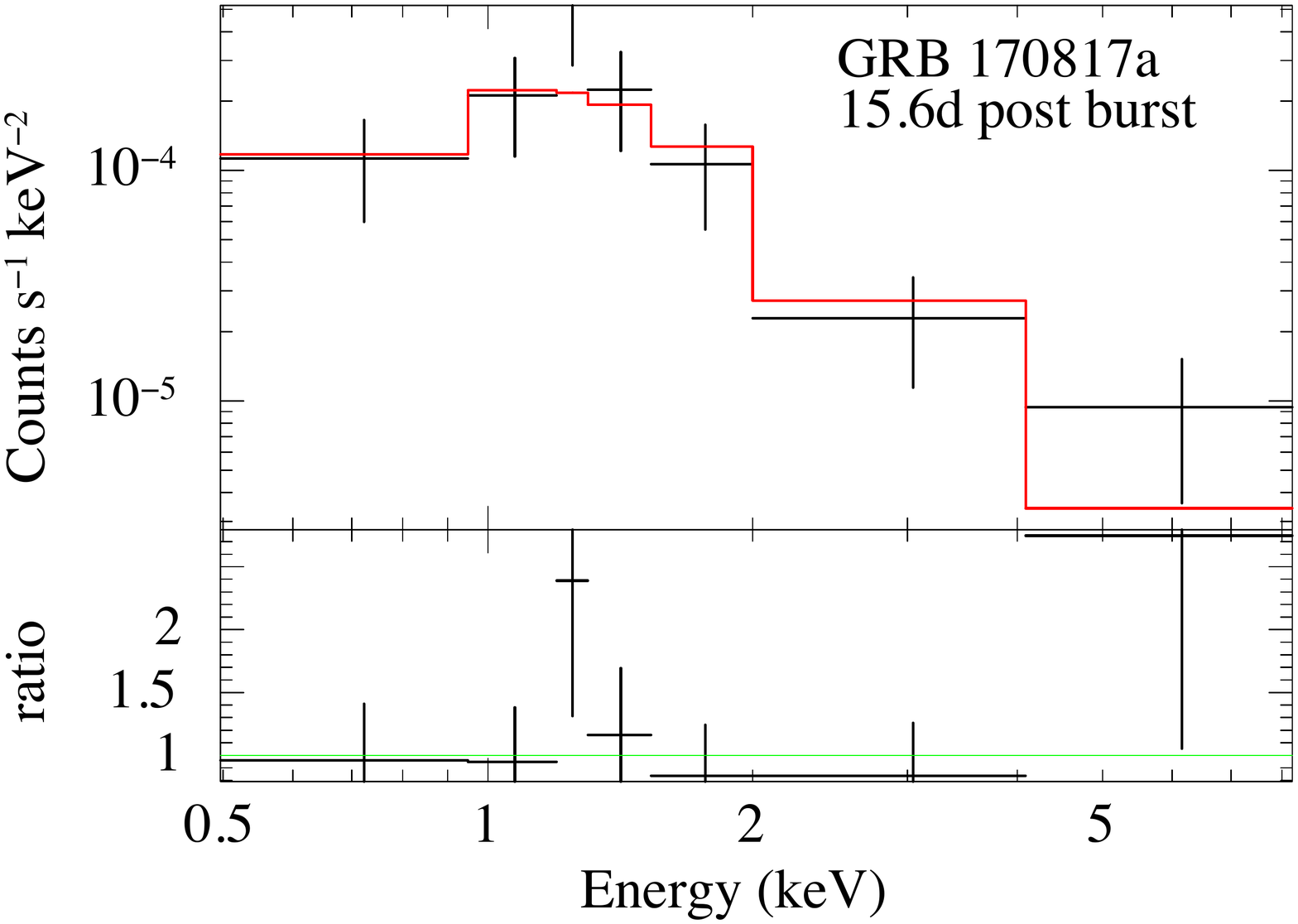}
\includegraphics[scale=0.32,angle=0]{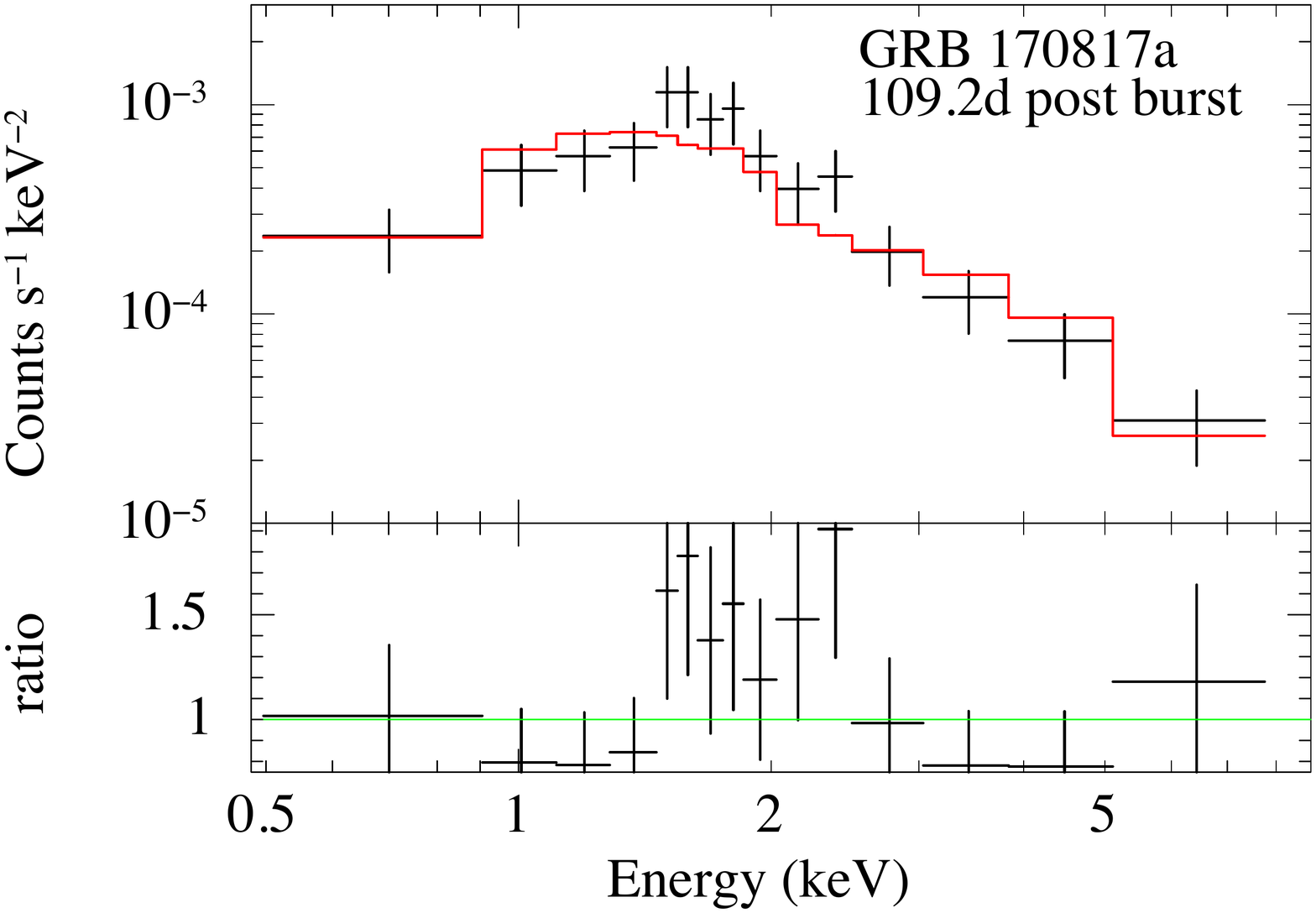}
}
\figcaption{ \Chandra\ co-added X-ray spectra from GRB170817A in a 93.4 ks exposure at 15.6 days (\textit{Left}) and a 98.8 ks exposure at 109.2 days post-burst (\textit{Right}). Spectral data and residuals are shown in black and the best-fit spectral models are in red. The neutral hydrogen absorption column is fixed to $N$\textsubscript{H}$=7.5 \times10^{20}$~cm$^{-2}$ (see Section \ref{sec:obs}). These spectra at 15.6 and 109.2 days each consist of two exposures that were jointly fit. In the joint fit for each epoch, the spectral index $\Gamma$ was tied between the exposures and the power-law normalization was left free. }
\label{fig:spec}
\end{figure*}
\begin{deluxetable*}{lccccccc}
\tablecaption{X-ray Source Properties at 15.6 days and 109.2 Days Post-Burst}
\tablehead{  
  \colhead{Source} & \colhead{Days} & \colhead{Count Rate} & \colhead{Power-Law} &  \colhead{Flux} & 
            \colhead{Luminosity~\tablenotemark{a,b}} & 
            \colhead{$\chi^2$/dof}\\
  \colhead{ID} & \colhead{Post-Burst} & \colhead{(${0.5-10}$ keV)} & \colhead{Index} & \colhead{(${0.3-8}$ keV)} & 
  			\colhead{(${0.3-10}$ keV)} &  \colhead{} \\
    \colhead{} & \colhead{} & \colhead{[$10^{-4}$~cts~s$^{-1}$]} & \colhead{$\Gamma$} & \colhead{[$10^{-14}~$erg~s$^{-1}$~cm$^{-2}$]} & \colhead{[$10^{38}$~erg~s$^{-1}$]} & \colhead{} &
  }
\decimals
\startdata
GRB170817A  &  15.6 & $3.5\pm0.7$ & {$2.4^{+0.8}_{-0.8}$} & $0.36^{+0.1}_{-0.07}$ & $10.4^{+2.0}_{-1.6}$  &  3.15/5\\
   &  109.2 & $14.7\pm1.3$ & {$1.62^{+0.27}_{-0.26}$} & $1.58^{+0.14}_{-0.13}$ & $42.5^{+3.7}_{-3.5}$ &  11.9/13
\vspace{3pt}\\
\hline
CXOU J130948 &  15.6 & $2.5\pm0.6$ & {$1.3^{+0.8}_{-0.8}$} & $0.4^{+0.1}_{-0.09}$   & $10.8^{+4.4}_{-2.4}$  & 1.77/3\\
   &  109.2 & $1.2\pm0.4$ & {$2.4\pm1.7$} & $0.2^{+0.3}_{-0.1}$   & $6.5^{+9.9}_{-4.7}$ &  0.26/2
\vspace{3pt}\\ 
\hline
CXOU J130946 & 15.6 & $3.3\pm0.7$ & {$-0.4^{+0.2}_{-0.8}$} & $1.1^{+0.1}_{-0.1}$ & $38.8^{+4.0}_{-9.8}$ &  3.68/5\\
   &  109.2 & $2.5\pm0.7$ & {$0.4\pm0.9$} & $0.5^{+0.5}_{-0.3}$ & $17.5^{+11.9}_{-10.3}$ &  2.41/3
\vspace{3pt}\\
\hline
NGC 4993 & 15.6 & $11.0\pm1.2$ & $1.5\pm0.4$ & $1.3\pm0.2$ & $34.8^{+3.2}_{-3.6}$  & 12.73/11\\
   & 109.2 & $12.3\pm1.2$ & {$1.4^{+0.4}_{-0.3}$} & $1.4^{+0.3}_{-0.2}$ & $38.9^{+5.8}_{-5.4}$ &  13.09/11
\vspace{3pt}\\
\enddata
\tablecomments{All reported errors represent $90\%$ confidence intervals. The neutral hydrogen absorption was frozen to $N$\textsubscript{H}$=7.5 \times10^{20}$~cm$^{-2}$ for all spectral fits, based on NGC 4993's A$_{\rm V}=0.338$ \citep{schlafly11}.}
\tablenotetext{a}{A luminosity distance of 42.5~Mpc was assumed for all sources.}
\tablenotetext{b}{The luminosities at 15.6 days listed here are larger than reported in \citet{haggard17} by a factor of 4, due to a previous calculation error.}
\label{tab:fluxes}
\end{deluxetable*}

X-ray observations of GW170817 provided important early indication that the associated sGRB GRB170817A was unlike any other GRB afterglow previously observed \citep{fong17}. \Chandra\ observations of the optical counterpart \citep[SSS17a;][]{coulter17} at $\sim$2 days post-burst yielded a non-detection \citep{margutti17}. However, \Chandra\ observations at $\sim$9 days post-burst revealed a new X-ray source coincident with the position of the optical transient \citep{troja17}. This delayed brightening of the X-ray counterpart has not been observed in standard GRB afterglows, which instead display consistent dimming in their X-ray light curves on timescales of days. The discovery of an X-ray counterpart was confirmed in additional \Chandra\ observations at $\sim$15 and 16 days post-burst, which surprisingly showed that the source had not dimmed significantly \citep{haggard17}. Unfortunately, Sun-constraints prevented further X-ray monitoring of GRB170817A after the last detection at 16 days post-burst, until early December 2017 ($\sim$109 days post-burst).

The origin of the observed X-ray emission from GRB170817A is still unclear. \citet{haggard17}, \citet{margutti17}, and \citet{troja17} all showed that the \Chandra\ X-ray light curve is consistent with the fading of a standard sGRB afterglow in which a simple top-hat jet is off-axis from the line of sight by 20-30$^\circ$. This interpretation is ostensibly supported by the low fluence observed in the prompt $\gamma$-ray emission \citep[e.g.,][]{abbott17c, goldstein17, murguia17}, which was a factor of $\sim$10$^3$ lower than previously observed in any other sGRBs. Modeling of the radio light curves of GRB170817A similarly showed consistency with off-axis sGRB afterglows \citep{alexander17, hallinan17}. If confirmed, this also makes GRB170817A the first off-axis sGRB ever observed, but other interpretations for the X-ray emission are not ruled out.

Alternatively, is it possible that the X-ray emission from GRB170817A is from the afterglow of a mildly-relativistic cocoon. Hydrodynamic simulations of GRB jets have shown that as the jet travels through the ejecta surrounding the burst, the jet head will shock the debris and produce a hot cocoon around the jet \citep{gottlieb18, lazzati17a, nakar17}. This cocoon can be mildly-relativistic, and shock the external medium after breakout to produce synchrotron afterglow emission in X-rays that is observable over a wide range of viewing-angles. Early evidence for a cocoon in GRB170817A was provided by the soft thermal tail observed in the $\gamma$-ray light curve of its prompt emission, which can be produced by cooling emission from a cocoon \citep{goldstein17}. \citet{piro17} also showed that the early blue emission of the optical transient associated with GRB170817A is consistent with shock cooling in a cocoon. 

More recently, \citet{mooley17} showed that the late-time radio light curve of GRB170817A rules out off-axis top-hat jet models, and instead supports an outflow afterglow interpretation for the origin of the radio emission. Continued radio monitoring of GRB170817A up to 107 days post-burst now reveals that the radio emission has continued to brighten monotonically, at a slower rate than predicted in simple jet models. \citet{mooley17} argue that the brightening radio light curve can be produced by a outflow in which the majority of the kinetic energy of the blast-wave is in the lower-velocity material. This leads to a continuous injection of energy into the shock, and thus a slow, monotonic rise of the afterglow emission. The outflow in these models could either be a mildly-relativistic cocoon that is shocked by the jet, or the high-velocity tail of the dynamical ejecta from the neutron star merger. Furthermore, an observable jet that breaks out of the debris is not required (or even favored).

Although the radio light curve of GRB170817A no longer favors off-axis top-hat jet models, more sophisticated structured jet models may still be able to describe the X-ray and radio light curves. \citet{lazzati17b} presented a simulation of a structured jet that breaks out of the post-merger debris. When observed off-axis, afterglow emission from material in the structured jet at progressively smaller angles from the jet axis become observable over time. \citet{lazzati17b} showed that this model predicts radio and X-ray properties that are a good match to the observations. The key differences between structured jet models and the outflow models of \citet{mooley17} are (1) the structured jet is not choked in the debris, and (2) the slow brightening of off-axis structured jet afterglow stems from the increasingly relativistic material that becomes observable over time.

The late-time X-ray light curve of GRB170817A can provide critical evidence to test our rapidly-evolving understanding of the EM transient. An X-ray light curve that brightens like the radio can confirm that the radio and X-ray emission have a common origin. This X-ray brightening would also provide evidence that the X-ray emission is the afterglow of an outflow, which could be a cocoon, dynamical ejecta, or a structured jet. Deep X-ray observations can also help determine the origin of the outflow by constraining its velocity, based on the time at which a synchrotron cooling break is observed to pass through the X-ray band.

In this Letter, we present deep \Chandra\ observations of GRB170817A at 109.2 days post-burst. Due to Sun constraints, these are the first X-ray observations since the previous \Chandra\ detection at 16 days post-burst. The outline of this Letter is as follows: In Section \ref{sec:obs}, we describe our \Chandra\ observations and data reduction procedure. In Section \ref{sec:discussion}, we discuss the origin of the X-ray and radio emission, and compare the observed X-ray light curve to predictions from outflow and structured jet afterglow models. We briefly summarize and conclude in Section \ref{sec:conclusion}. 

\section{X-ray Observations and Analysis}
\label{sec:obs}

New X-ray observations of GRB170817A were obtained via a \Chandra\ Director's Discretionary Time allocation (PI: Wilkes, Program Number 18408601). This program obtained two exposures of GRB170817A: (1) a 74.09 ks exposure (ObsID 20860) beginning at 2017 December 2.08 UT, approximately 108 days post-burst, and (2) a 24.74 ks exposure (ObsID 20861) beginning at 2017 December 6.45 UT, approximately 111 days post-burst. Both these exposures were acquired using \Chandra's ACIS-S3 chip in VFAINT mode. We use CIAO v.4.9 \citep[CALDB v4.7.6;][]{fruscione06} for reduction and analysis of these X-ray data. We reprocess the level 2 events files and use CIAO's {\tt repro} script to apply the latest calibrations. Since the two new exposures are close in time and the X-ray emission of GRB170817A is not expected to vary significantly over $\sim$4 day timescales, we co-add the two data sets into one 98.83~ks exposure at 109.2 days post-burst.  

Figure \ref{fig:image} shows our latest \Chandra\ 0.5--8 keV image of GRB170817A at 109.2 days post-burst ({\it Right}), in comparison to the \Chandra\ image from a similar exposure at 15.6 days post-burst from \citet{haggard17} ({\it Left}). The previously-detected X-ray source at the position of GRB170817A is still detected in this latest observation, along with the three other previously-detected X-ray sources in the field: CXOU~J130948, CXOU~130946, and the host galaxy NGC~4993. We construct a difference image by subtracting the 15.6 day image from the 109.2 day image, scaled by their exposure times. This difference image (Figure \ref{fig:image}, {\it Right}) shows that GRB170817A has brightened in the latest observations.

To measure the flux of each of the four sources, we first apply the point-source detection algorithm {\tt wavdetect} to each $0.5-7$~keV image to determine the centroid positions of each source. Following the same procedure from \citet{haggard17}, we extract spectra from the point sources GRB170817A, CXOU~J130948, and CXOU 130946 using  extraction regions with radii 1\secspt97 (corresponding to $\sim$90\% encircled energy fraction near the \Chandra\ on-axis position). For the host galaxy NGC~4993 we adopt an extraction region with a $2$\secspt$95$ radius, large enough to include most of the galaxy's X-ray flux while avoiding contamination from nearby CXOU~J130948. These regions are shown in Figure \ref{fig:image}. We obtain background photons from a large region on the same chip that does not overlap other sources. 

We extract spectra and response files for the four detected X-ray sources using the CIAO {\tt specextract} tool from the individual observations.  The resultant files are then co-added to improve statistics.  We fit the combined spectra using XSPEC v12.9.0 \citep{arnaud96}, with atomic cross sections from \citet{verner96} and abundances from \citet{wilms00}. For each of the four sources, we assume absorbed power-law spectral models with fixed $N_\mathrm{H} = 7.5\times10^{20}$ ~cm$^{-2}$. The best-fit power-law spectral indices, count rates, absorbed $0.3-8$~keV fluxes, and $0.3-10$~keV luminosities are listed in Table \ref{tab:fluxes}. 

Our X-ray analysis confirms that the X-ray flux of GRB170817A has brightened significantly, with an increase in detection significance from $7.2\sigma$ at 15.6 days to $17.0\sigma$ at 109.2 days post-burst.  We detect a $0.5-8$~keV source count rate of $14.7\pm1.3$~counts~s$^{-1}$ at 109.2 days, a factor of $\sim$4 larger than the previous detection at 15.6 days. This count rate at 109.2 days corresponds to an absorbed flux of $F_\mathrm{0.3-8~keV} = 15.8 \times 10^{-15}$ erg s$^{-1}$ cm$^{-2}$, and an unabsorbed luminosity of $L_\mathrm{0.3-10~keV} = 4.3 \times10^{39}$~erg~s$^{-1}$. 

The extracted X-ray spectrum of GRB170817A at 109.2 days is shown in Figure \ref{fig:spec}, along with our previous spectrum at 15.6 days from \citet{haggard17} for comparison. The spectrum is well-described by an absorbed power-law model with $\chi^2_{\nu}=0.92$. We note a possible excess at energies near $1.5-2.0$ keV, which could be due to line emission from Si or S, characteristic of supernova remnants observed with \Chandra. We test for additional model components by adding Gaussians at various line centers to the absorbed power-law model. Though adding a line component, e.g., at $E=1.8$~keV, can lower the reduced $\chi^2$ to $0.49$ and remove the excess around $2$~keV, the Gaussian does not dominate the spectrum. The spectral index is also reduced, but its larger uncertainty ($\Gamma=1.45^{+0.45}_{-0.36}$) places the value within the error interval of the power-law-only fit. Hence, inclusion of an emission line is not yet statistically supported by the \Chandra\ data. If GRB170817A continues to brighten we may be able to better quantify this tantalizing line emission. We also note that a simple absorbed thermal model ({\tt blackbody}, $kT=0.63\pm0.09$) can be fit to the spectrum with a similar reduced $\chi^2$. However, none of the current models for the late-time X-ray emission predict a thermal spectrum, and the physical process that links the brightening radio and X-ray emission would be unclear.

The flux and spectrum of the host-galaxy NGC 4993 and CXOU J130946 are consistent with our previous deep \Chandra\ observations. CXOU J130948 is known to be variable in X-ray, and while our spectral analysis shows its spectrum is consistent with those previously reported, the best-fit parameters are not well-constrained. This source is visibly dimmer at 109.2 days (Figure \ref{fig:image}) and shows a $\sim48\%$ decrease in count rate between the two epochs \citep[Table \ref{tab:fluxes};][]{margutti17,haggard17}. 

\begin{figure}[t!]
\center{
\includegraphics[scale=0.58]{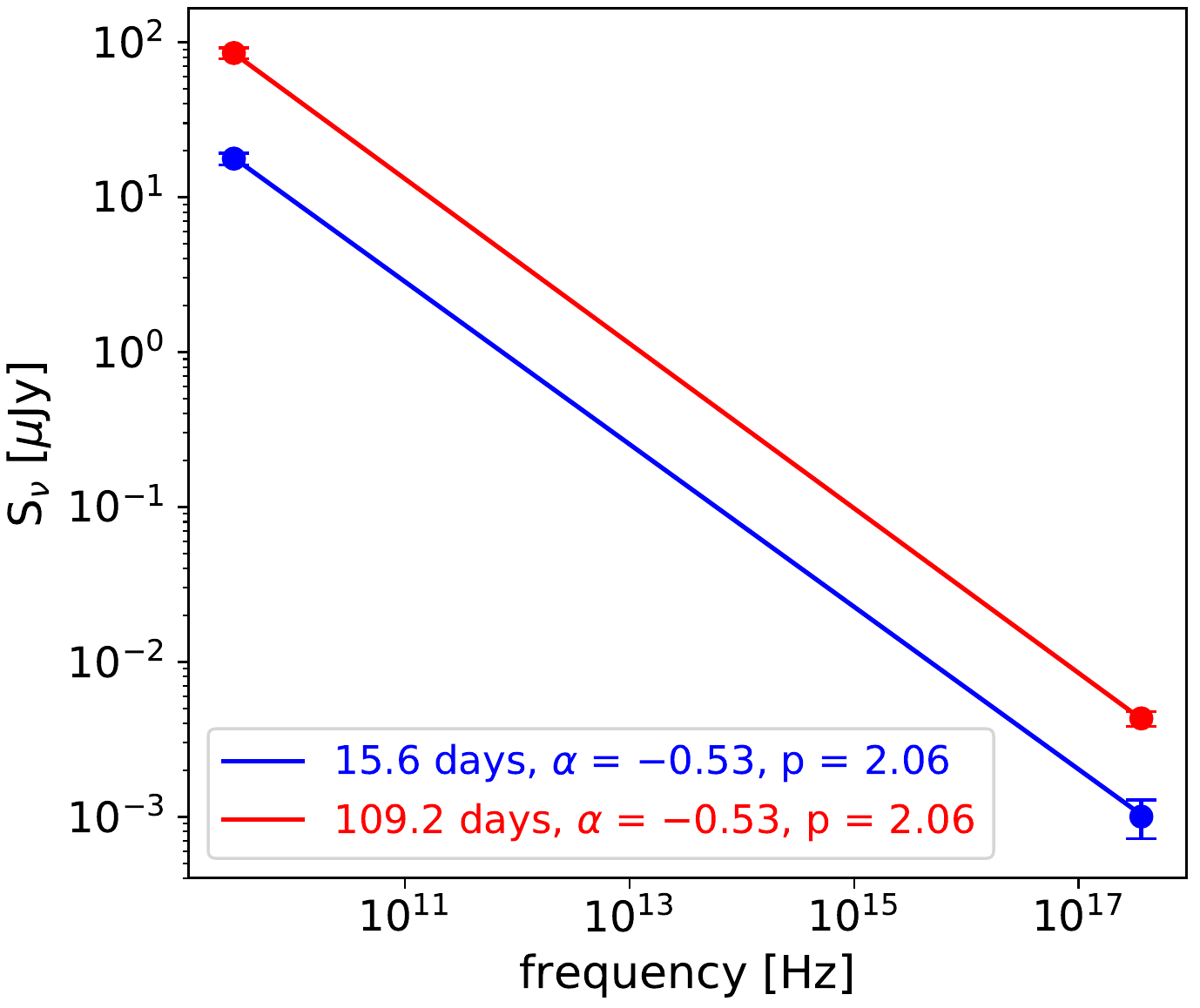}
}
\figcaption{Comparison of the radio to X-ray flux densities at 15.6 days (blue points) and 109.2 days post-burst (red points). The X-ray-to-radio spectral index $\alpha$ remains constant at $-0.53$ over this timespan, indicating that the X-ray emission brightened at the same rate as the radio.}
\label{fig:SED}
\end{figure}

\section{Discussion}
\label{sec:discussion}

The observed X-ray brightening of GRB170817A at 109.2 days in comparison to previous X-ray observations at 9 and 15.6 days has profound implications for our understanding of the EM transient. In Section \ref{ssc:radio} below, we first test whether the X-ray and radio emission have the same origin, by comparing the X-ray fluxes to the radio observations. We find that the X-rays brightened at a rate similar to the radio emission, confirming that they share a common origin. In Section \ref{ssc:outflow}, we consider the possibility that both the X-ray and radio emission are due to the afterglow of an outflow, either from a cocoon, dynamical ejecta, or a structured jet. We find that the X-ray light curve is well-described by predictions from outflow models based on radio observations.

\begin{figure*}[t!]
\center{
\includegraphics[scale=0.75]{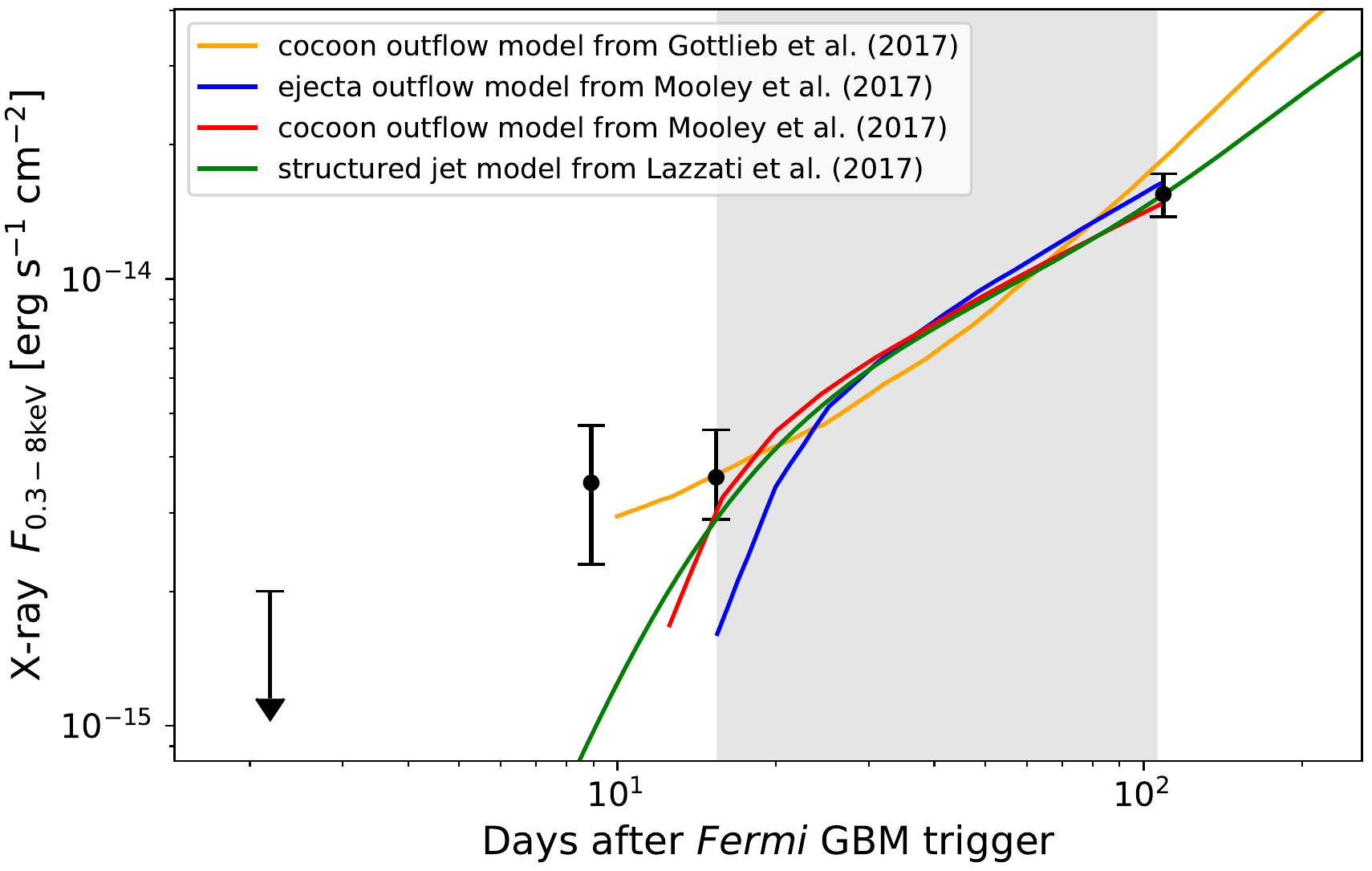}
}
\figcaption{\Chandra\ X-ray light curve of GRB170817A (black points), including our new observations at 109.2 days post-burst. The predicted X-ray light curves from the cocoon model of \citet{gottlieb17} (solid orange line), the cocoon (solid red line) and ejecta (solid blue line) outflow models of \citet{mooley17}, and the structured jet model of \citet{lazzati17b} are also shown. The gray shaded region is the timespan over which \Chandra\ observations were not possible due to Sun constraints. All uncertainties shown are 90\% confidence intervals. The brightening X-ray light curve is well-described by cocoon, dynamical ejecta, or structured jet models, although several of these models under-predict the detected early X-ray flux at 9 days post-burst.}
\label{fig:lightcurve}
\end{figure*}


\subsection{Comparison of X-ray and Radio Emission: \\ Evidence for A Common Origin}
\label{ssc:radio}

The similar brightening of both X-ray and radio emission from GRB170817A suggests that the X-ray and radio emission may have a common origin. This can be confirmed if the X-ray-to-radio spectral index $\alpha$ (where $S_\nu \sim \nu^\alpha$) is measured to be constant between 15.6 and 109.2 days post-burst. In this scenario, the X-ray light curve would be consistent with a scenario in which afterglow emission from an outflow is the origin of both the radio and X-ray emission, while off-axis GRB afterglows from simple top-hat jet models are ruled out. If instead $\alpha$ is measured to have steepened, this can either indicate that the X-ray and radio emission have different origins, or that they have the same synchrotron origin but a cooling break has passed through the X-ray band.

We compare the X-ray-to-radio spectral index $\alpha$ at 15.6 and 109.2 days and find that it remains constant, confirming a common origin for both the X-ray and radio emission. For the X-ray, we use the measured X-ray flux densities at 15.6 days from \citet{haggard17} and at 109.2 days reported here. For the radio, we use the VLA 3 GHz radio light curve from \citet{hallinan17} and \citet{mooley17}, which covers a timespan from 16.4 to 93.1 days post-burst. \citet{mooley17} showed that this radio light curve is well-fit by a power-law with slope $t^{0.8}$, and thus we also fit a power-law to extrapolate the radio flux density at 15.6 and 109.2 days. The uncertainties on these two extrapolated radio flux densities are estimated through Monte Carlo re-sampling of the radio flux density measurements in the light curve based on their uncertainties, then re-performing the fit and extrapolation. Figure \ref{fig:SED} compares the X-ray and radio flux densities at the two epochs. The fitted X-ray-to-radio spectral index is $-0.53 \pm 0.02$ at 15.6 days, and $-0.53 \pm 0.01$ at 109.2 days. Thus, we detect no change in $\alpha$, implying that the X-rays brightened at the same rate as the radio. The similar rate of brightening in the X-ray and radio light curves thus confirms a common origin. Moreover, the consistency between the radio spectral index \citep[$\alpha = 0.60$;][]{alexander17, mooley17} and X-ray photon index ($\Gamma = 1.62$, where $\alpha = \Gamma - 1$; see Table \ref{tab:fluxes}) further supports this conclusion.

\subsection{Comparison with Models}
\label{ssc:outflow}

\citet{mooley17} showed that the brightening radio light curve of GRB170817A rules out simple top-hat off-axis jet models, and is instead well-described by models of afterglow emission from an outflow. This outflow can be a mildly-relativistic cocoon shocked by the jet head, or the high-velocity tail of dynamical ejecta from the neutron star coalescence. In both cases, the slow and monotonic rise of the radio light curve implies that the blast-wave must have a continued injection of kinetic energy. \citet{mooley17} show that both cocoon and dynamical ejecta models in which the majority of the kinetic energy is in the lower velocity material provide excellent fits to the radio light curve. These models also predict that the X-ray light curve will rise at the same rate as the radio, which we can directly test with our X-ray light curve.

We produce predicted X-ray light curves based on the radio light curves from both the cocoon and ejecta outflow models presented in \citet{mooley17}, as well as the cocoon outflow model from \citet{gottlieb17}. The X-ray emission is estimated by scaling the model radio light curves to the X-rays based on the X-ray-to-radio flux ratio at 15.6 days post-burst. This assumes that the synchrotron cooling break has not yet passed through the X-rays, and thus the X-ray-to-radio spectral index remains constant. This assumption is justified because the fiducial models used by \citet{mooley17} to fit the radio light curve predict that the synchrotron cooling frequency is still well above the \Chandra\ band at 109.2 days post-burst. The predicted X-ray light curves from these outflow models are shown in Figure \ref{fig:lightcurve}. 

The X-ray light curves predicted by the three cocoon and dynamical ejecta outflow models in Figure \ref{fig:lightcurve} are a good match to our observed X-ray flux at 109.2 days post-burst. Although the cocoon and dynamical ejecta outflow models of \citet{mooley17} under-predict the early-time X-ray flux at 9 days, they are fitted to the radio light curve (which do not cover such early times) rather than the X-ray light curve. A more complete fit of these models that also incorporates the X-ray data will likely improve the fit. The general agreement of the X-ray light curve with these models is consistent with a scenario in which the afterglow of a cocoon or dynamical ejecta produce both the X-ray and radio emission, and implies that the synchrotron cooling break has not yet passed through the X-ray band.

If the synchrotron cooling frequency is still above the X-ray band, continued deep \Chandra\ observations will provide the first indication of the cooling break, via a change in the X-ray photon index $\Gamma$. Across the break frequency, the X-ray photon index should be observed to steepen by $\Delta$$\Gamma = 0.5$ (where $\alpha = \Gamma - 1$), from our measured $\Gamma = 1.62 \pm 0.27$ at 109.2 days. Since the synchrotron cooling frequency is strongly dependent on the velocity of the outflow, the detection of a cooling break in X-ray monitoring can provide powerful constraints on critical parameters in these models. We thus strongly encourage additional deep \Chandra\ X-ray observations of GRB170817A toward this end.

Although \citet{mooley17} showed that the radio light curve rules out afterglow emission from off-axis top-hat jet models, more sophisticated off-axis structured jets can still be supported by the current data. For example, \citet{lazzati17b} presented a simulation of an off-axis structured jet, and showed that the afterglow emission can produce the observed X-ray and radio properties. When observed off-axis, the afterglow emission from structured jets will come from increasingly more relativistic material closer to the jet axis as the effects of beaming become less severe over time. This can produce a slower brightening in its afterglow light curve in comparison to off-axis top-hat jets, akin to the observed slowly-brightening radio and X-ray light curves of GRB170817A. Figure \ref{fig:lightcurve} shows that the predicted X-ray light curve from this off-axis structured jet simulation of \citet{lazzati17b} is also well-matched to our observed X-ray flux at late-times. Since it is currently difficult to robustly differentiate whether the origin of the radio and X-ray emission is from the afterglow of a cocoon, dynamical ejecta, or a structured jet, continued X-ray and radio monitoring of GRB170817A should be avidly pursued.

\section{Conclusion}
\label{sec:conclusion}

We present late-time \Chandra\ observations of neutron star coalescence GW170817/GRB170817A at 109.2 days post-burst, the first sensitive X-ray observations possible since a previous detection at 15.6 days. These data show that the X-ray counterpart has brightened at the same rate as its radio light curve. We show that the X-ray light curve is an good match to predictions from outflow models, in which the outflow is a cocoon, dynamical ejecta, or a structured jet. Our observations thus support a scenario in which both the X-ray and radio emission are the afterglow of an outflow, although the exact origin of the outflow is still uncertain. Finally, the X-ray brightening strengthens the argument that simple top-hat jet models are not consistent with the latest observations.

Continued \Chandra\ monitoring of GRB170817A will be critical for validating outflow models. Since the synchrotron cooling frequency is still above the X-ray band, additional deep observations by \Chandra\ will provide the first indication of the cooling break by detecting a steepening in the X-ray photon index. A detection of the cooling break will provide powerful constraints on the physical parameters of the outflow producing the X-ray and radio emission. Our observations of GW170817/GRB170817A presented here highlight the unique role of \Chandra\ in opening the multi-messenger gravitational wave era. 

\acknowledgments
The authors thank Belinda Wilkes and the \Chandra\ scheduling, data processing, and archive teams for making these observations possible. This work was supported by \Chandra\ Award Number GO7-18033X, 
issued by the {\it Chandra X-ray Observatory Center}, which is operated by the Smithsonian Astrophysical Observatory for and on behalf of the National Aeronautics Space Administration (NASA) under contract NAS8-03060. 
J.J.R., M.N., and D.H. acknowledge support from a Natural Sciences and Engineering Research Council of Canada (NSERC) Discovery Grant and a Fonds de recherche du Qu\'{e}bec--Nature et Technologies (FRQNT) Nouveaux Chercheurs Grant. J.J.R. and M.N. acknowledge funding from the McGill Trottier Chair in Astrophysics and Cosmology. D.H. acknowledges support from the Canadian Institute for Advanced Research (CIFAR). PAE acknowledges UKSA support. 

\facility{CXO}

\bibliographystyle{apj}

\end{document}